\documentclass[prb,twocolumn,showpacs,preprintnumbers,amsmath,amssymb]{revtex4}

\usepackage{graphicx}
\usepackage{dcolumn}
\usepackage{bm}

\newcommand{\mapright}[1]{\smash{\mathop{\hbox to 1.0cm{\rightarrowfill}}\limits^{#1}}}

\begin{document}

\preprint{}

\title{Theory of Macroscopic Quantum Tunneling in High-$T_c$ $c$-Axis Josephson Junctions}

\author{Takehito Yokoyama,$^1$ Shiro Kawabata,$^2$ Takeo Kato,$^3$ and Yukio Tanaka$^1$}
\affiliation{$^1$Department of Applied Physics, Nagoya University, Nagoya, Aichi, 464-8603, Japan%
\\
and CREST, Japan Science and Technology Corporation (JST), Kawaguchi, Saitama 332-0012,
Japan \\
$^2$Nanotechnology Research Institute (NRI), National Institute of Advanced Industrial Science and Technology (AIST), Tsukuba, Ibaraki, 305-8568, Japan  and CREST, JST, Kawaguchi, Saitama 332-0012, Japan \\
$^3$The Institute for Solid State Physics (ISSP), University of Tokyo,
Kashiwa, Chiba, 277-8581, Japan}
\date{\today}

\begin{abstract}
We study macroscopic quantum tunneling (MQT) in $c$-axis twist Josephson junctions made of high-$T_c$ superconductors in order to clarify the influence of the anisotropic order parameter symmetry (OPS) on MQT.
The dependence of the MQT rate on the twist angle $\gamma$ about the $c$-axis is calculated by using the functional integral and the bounce method.
Due to the  $d$-wave OPS, the $\gamma$ dependence of  standard deviation of the switching current distribution and the crossover temperature from thermal activation to MQT are found to be given by $\cos2\gamma$ and $\sqrt{\cos2\gamma}$, respectively.
We also show that a dissipative effect resulting from the nodal quasiparticle excitation on MQT is negligibly small, which is consistent with recent MQT experiments using Bi${}_2$Sr${}_2$CaCu${}_2$O${}_{8 + \delta}$ intrinsic junctions.
These results indicate that MQT in $c$-axis twist junctions becomes a useful experimental tool for testing the OPS of high-$T_c$ materials at low temperature, and suggest high potential of such junctions for qubit applications.
\end{abstract}

\maketitle

\section{Introduction}
The phenomenon of macroscopic quantum tunneling (MQT) has attracted much attention of experimentalists and theorists for many years. \cite{rf:MQT1,rf:MQT2,rf:MQT3}  
Among several works on MQT, Josephson junctions have been intensively studied.
In the current biased Josephson junctions, the states of non-zero supercurrent can move to lower-lying minima of the potential through the potential barrier by MQT. 
Ivanchenko and Zi'lberman showed the possibility of observing MQT in such systems.~\cite{rf:Ivanchenko68}
As was predicted by Caldeira and Leggett, MQT is suppressed by the dissipation effect.~\cite{rf:CL81,rf:CL83}
Later, MQT and the dissipative effects on MQT were experimentally observed in $s$-wave Josephson junctions.~\cite{rf:Voss81,rf:Jackel81,rf:Cleland88,rf:Devoret92}

Recently, the MQT theories for $s$-wave Josephson junctions~\cite{rf:CL81,rf:CL83,Ambegaokar,Eckern,Esteve,Zaikin} have been extended to $d$-wave systems.~\cite{Kawabata1,Kawabata2,Kawabata3,Kawabata4,Kawabata5,Kawabata6-1,Kawabata6-2} 
It was claimed that the influence of  the quasiparticle excitation on MQT is negligible despite of the existence of the line nodes which result from the $d$-wave order parameter symmetry (OPS).~\cite{Harlingen,Tsuei}
Therefore, in  $c$-axis Josephson junctions, e.g., intrinsic junctions~\cite{Yurgens} or cross-whisker junctions,~\cite{Klemm,Takano1,Badica} the crossover temperature $T^*$ from thermal activation (TA) to MQT was predicted to be quite high.
This can be ascribed not only to the weak dissipative nature~\cite{Kawabata1,Kawabata3,Kawabata4} but also the large Josephson plasma frequency $\omega_p$ of $d$-wave junctions, i.e., $T^* \propto \omega_p$.~\cite{rf:MQT1,rf:MQT2}
On the other hand, in the case of in-plane $d$-wave junctions which are parallel to the CuO${}_2$ plane, e.g., grain boundary junctions~\cite{Hilgenkamp,Tafuri1}  or ramp-edge junctions,~\cite{Arie} it was found that the zero energy Andreev bound states (ZES)~\cite{Hu,TK95,KashiwayaTanaka,Lofwander} give the Ohmic dissipative effect on MQT.~\cite{Kawabata2,Kawabata4,Kawabata5} 
Later, the above MQT theories have been extended to explore macroscopic quantum coherence~\cite{Khveshchenko1,Khveshchenko2,Umeki} and propose a $d$-wave phase qubits.~\cite{Kawabata7}

The TA-related phenomena in $d$-wave junctions were already observed experimentally by many groups.~\cite{TA0,TA1,TA2,TA3,TA4,TA5,TA6,TA7,TA8,TA9}  
Until recently, however, no experiments of MQT have been reported.
First successful observations of MQT in $d$-wave junctions were performed by Bauch $et$ $al.$~\cite{Bauch1} and Inomata $et$ $al.$,~\cite{Inomata} using YBCO bi-epitaxial grain boundary junctions,~\cite{Tafuri1,Lombardi1,Tafuri2,Lombardi2,Bauch3} and high-quality Bi${}_2$Sr${}_2$CaCu${}_2$O${}_{8 + \delta}$ (Bi2212) intrinsic junctions, respectively.
Subsequently, several groups have observed MQT\cite{Jin,Matsumoto,Kashiwaya,Kashiwaya2,Li} and micro-wave assisted MQT~\cite{Bauch2,Jin,Inomata2,Inomata3} in such systems.
They reported that $T^*$ of $c$-axis (Bi2212 intrinsic) junctions is high (0.5$\sim$1K)~\cite{Inomata,Jin,Matsumoto,Kashiwaya} compared with the high-quality $s$-wave junction in which $T^*$ is at most 0.3K.~\cite{Wallraff}
This result is consistent with theoretical predictions.~\cite{Kawabata1,Kawabata3,Kawabata4}

In the previous MQT studies for $d$-wave junctions, the advantage of a large gap value of $d$-wave superconductors has been mainly emphasized.
On the other hand, in $d$-wave junctions,  there appears a new degree of freedom, i.e., a directional dependence of the anisotropic order parameter. 
 This directional dependence produces many intriguing phenomena.\cite{Harlingen,Tsuei,KashiwayaTanaka,Lofwander}
  Additionally, angular dependence of the Josephson current in $d$-wave junctions explicitly reflects the $d$-wave OPS.~\cite{SR,Yokoyama,Takano1} 
 Therefore, how  MQT depends on the relative angle of the lobe directions between two order parameters is an interesting problem.

In the present paper, we study the dependence of the MQT rate (the inverse lifetime of the metastable state) on the twist angle $\gamma$ for the $c$-axis twist Josephson junction (see Fig. 1) by use of the functional integral and the bounce method. 
We also investigate favorable conditions for the realization of  MQT at high crossover temperature $T^*$. 
Recently, it was found that intrinsic junction stacks exhibit a gigantic enhancement of the MQT rate.~\cite{Jin,Machida1,Machida2,Fistul1,Fistul2,Nori}
 In the present paper, however, we treat a single junction in order to clarify the effect of the anisotropic OPS which is peculiar to high-$T_c$ superconductors. 
Note that such single intrinsic junctions made of Bi2212 single crystals have been recently fabricated by use of the Ar-ion etching method.~\cite{You1,You2} 
We will also present supplementary explanations and results of the previous papers~\cite{Kawabata1,Kawabata3,Kawabata4} in which the quasiparticle dissipation effect on MQT is mainly discussed.

The organization of the present paper is as follows. 
In section II, we derive the effective action for $d$-wave junctions and formulate the theory for the calculation of $T^*$. 
In section  III, we present calculated results of $T^*$. 
The advantages of $c$-axis twist Josephson junctions for qubits applications are briefly discussed in section IV.
In section V, a summary of the results in the present paper is given. 

\begin{figure}[tb]
\begin{center}
\includegraphics[width=8.5cm]{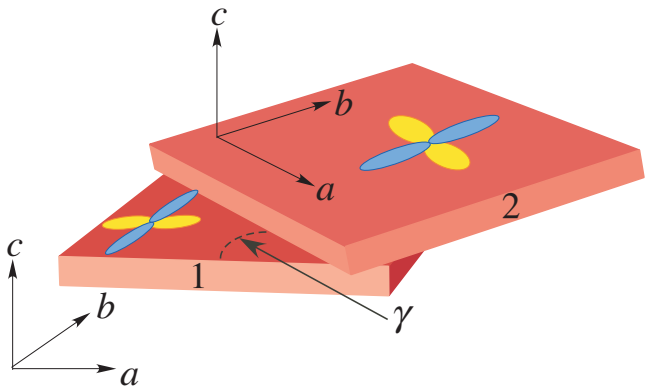}
\end{center}
\caption{(color online) Schematic picture of the $c$-axis twist Josephson junction. $\gamma$ is the twist angle around the $c$-axis. }
\label{f1}
\end{figure}

%
%
%
%
\section{Formulations}
\subsection{Effective action for $d$-wave junctions}

Let us start from a microscopic model of the $d$-wave superconductor/insulator/$d$-wave superconductor Josephson junction, described by the grand canonical Hamiltonian,
\begin{eqnarray}
{\cal H} = {\cal H}_1 +{\cal H}_2 +{\cal H}_T +{\cal H}_Q
\label{hamiltonian}
,
 \end{eqnarray}
where ${\cal H}_1$ and  ${\cal H}_2$ are Hamiltonians for the $d$-wave superconductors 1 and 2:
\begin{eqnarray}
{\cal H}_{1(2)}
&= &
\sum_\sigma \int d \mbox{\boldmath $r$} 
\ 
\psi_{1(2) \sigma}^\dagger \left( \mbox{\boldmath $r$}  \right)
\left( - \frac{\hbar^2 \nabla^2 }{2 m}  - \mu \right)
\psi_{1(2) \sigma}\left( \mbox{\boldmath $r$}  \right)
\nonumber\\
&-&
\frac{1}{2} \sum_{\sigma,\sigma'}  
\int d \mbox{\boldmath $r$}   d \mbox{\boldmath $r$}'
\psi_{1(2) \sigma}^\dagger \left( \mbox{\boldmath $r$}  \right)
\psi_{1(2) \sigma'}^\dagger \left( \mbox{\boldmath $r$}'  \right)
\nonumber\\
&\times&
g_{1(2)} \left( \mbox{\boldmath $r$}  - \mbox{\boldmath $r$}' \right)
\psi_{1(2) \sigma'} \left( \mbox{\boldmath $r$}'  \right)
 \psi_{1(2) \sigma} \left( \mbox{\boldmath $r$}  \right)
 ,
 \end{eqnarray}
where $m$ is the electron mass, $\mu$ is the chemical potential and $\psi$ ($\psi^\dagger$) is the fermion field operator.
In order to obtain the anisotropic order parameter, the anisotropic attractive interaction $g \left( \mbox{\boldmath $r$}  - \mbox{\boldmath $r$}' \right)$ has to be taken into account. 
The third term in Hamiltonian (\ref{hamiltonian}), i.e., 
\begin{eqnarray}
{\cal H}_T
= 
\sum_\sigma \int d \mbox{\boldmath $r$} d \mbox{\boldmath $r$}'
\ 
\left[
t \left( \mbox{\boldmath $r$},\mbox{\boldmath $r$}' \right) 
\psi_{1 \sigma}^\dagger \left( \mbox{\boldmath $r$}  \right)
\psi_{2 \sigma} \left( \mbox{\boldmath $r$}' \right)
+
\mbox{h.c.}
\right]
 \end{eqnarray}
describes the tunneling of electrons between the two sides of the junctions,
and 
\begin{eqnarray}
{\cal H}_Q
= 
\frac{\left( Q_1 - Q_2 \right)^2}{8C}
 \end{eqnarray}
is the charging Hamiltonian
where $C$ is the capacitance of the junction and $Q_{1(2)}$ is the operator for the charge on the superconductor 1 (2), which can be written as 
\begin{eqnarray}
Q_{1(2)}
= 
e \sum_\sigma \int d \mbox{\boldmath $r$}
\psi_{1(2) \sigma}^\dagger \left( \mbox{\boldmath $r$}  \right)
\psi_{1(2) \sigma} \left( \mbox{\boldmath $r$}  \right).
 \end{eqnarray}

The procedure to derive the effective action is the same as $s$-wave junctions.~\cite{Eckern,Ambegaokar,Zaikin,Awaka} 
By using the functional integral method,~\cite{rf:Popov,rf:Nagaosa} the ground partition function ${\cal Z}$ for the system can be written as 
\begin{eqnarray}
{\cal Z}
=
\int  {\cal D} \bar{\psi}_1{\cal D} \psi_1 {\cal D} \bar{\psi}_2 {\cal D} \psi _2
\exp
\left[
  - \frac{1}{\hbar} \int_{0}^{\hbar \beta} d \tau 
    {\cal L} (\tau) 
\right]
,
\end{eqnarray}
where $\beta=1/k_B T$, $ \psi(\bar{\psi})$ is the Grassmann field which corresponds to the fermionic field operator $\psi (\psi^\dagger)$, and the Lagrangian ${\cal L}$ is given by
\begin{eqnarray}
	  {\cal L} (\tau) 
	  =
	  \sum_\sigma \sum_{i=1,2} \int d \mbox{\boldmath $r$} 
	 \bar{\psi}_{i \sigma} \left( \mbox{\boldmath $r$}  , \tau \right) 
    \partial_\tau 
	\psi _{i \sigma} \left( \mbox{\boldmath $r$}  , \tau \right) 
	 + {\cal H} (\tau) 
	 .
\end{eqnarray}
We use the Hubbard-Stratonovich transformation~\cite{Kleinert,Sharapov}
\begin{eqnarray}
e^{ 
  -\frac{1}{\hbar} 
     \int_{0}^{\hbar \beta} d \tau
    \int d \mbox{\boldmath $r$} d \mbox{\boldmath $r$}'
    \bar{\psi}_{\uparrow} \left(   \mbox{\boldmath $r$}' , \tau \right)
    \bar{\psi}_{\downarrow} \left(   \mbox{\boldmath $r$} , \tau \right)
  g \left(  \mbox{\boldmath $r$} - \mbox{\boldmath $r$}'  \right)
    \psi_{\downarrow} \left(   \mbox{\boldmath $r$} , \tau \right)
    \psi_{\uparrow} \left(   \mbox{\boldmath $r$}', \tau \right)
}
\nonumber\\
=
\int {\cal D} \bar{\Phi}  ( \mbox{\boldmath $r$} , \mbox{\boldmath $r$}'  ,\tau)
 {\cal D} \Phi ( \mbox{\boldmath $r$} , \mbox{\boldmath $r$}'  ,\tau)
\exp \left[
\frac{1}{\hbar} 
 \int_0^{\hbar \beta} d \tau
 \int d \mbox{\boldmath $r$} d \mbox{\boldmath $r$}'
 \right.
 \nonumber\\
 \left.
 \times
\left\{
    -
	\frac{\left|  \Phi ( \mbox{\boldmath $r$} , \mbox{\boldmath $r$}'  ,\tau) \right|^2}
	{ g \left(  \mbox{\boldmath $r$} - \mbox{\boldmath $r$}'  \right)}
	+
	\bar{\Phi}  ( \mbox{\boldmath $r$} , \mbox{\boldmath $r$}'  ,\tau)
	\psi_{\downarrow} \left(   \mbox{\boldmath $r$} , \tau \right)
    \psi_{\uparrow} \left(   \mbox{\boldmath $r$}', \tau \right)
\right.
   \right.
   \nonumber\\
   \biggr.
   \biggr.
	+
   \bar{\psi}_{\uparrow} \left(   \mbox{\boldmath $r$} , \tau \right)
    \bar{\psi}_{\downarrow} \left(   \mbox{\boldmath $r$}' , \tau \right)
	\Phi  ( \mbox{\boldmath $r$} , \mbox{\boldmath $r$}'  ,\tau)
     \biggr\}
	 \biggr]
	 ,
\end{eqnarray}
in order to remove the term $\psi^4$ in the Hamiltonian ${\cal H}( \tau)$.
This introduces a complex order parameter field $\Phi  ( \mbox{\boldmath $r$} , \mbox{\boldmath $r$}'  ;\tau)$.
The resulting action is only quadratic in the Grassmann field, so that the functional integral over this number can readily be performed explicitly.

Next,  we perform the variable transformation (or gauge transformation)\cite{Ambegaokar,Eckern,Zaikin}
\begin{eqnarray}
\Phi \left(\mbox{\boldmath $r$}_a , \mbox{\boldmath $r$}_b ,\tau  \right) &=& \Delta \left( {{\bf{R}},{\bf{r}},\tau } \right)\exp \left[ {i\phi \left( {{\bf{R}},{\bf{r}},\tau } \right)} \right] ,
\\
\bar{\Phi} \left( \mbox{\boldmath $r$}_a , \mbox{\boldmath $r$}_b ,\tau  \right)& =& \Delta \left( {{\bf{R}},{\bf{r}},\tau } \right)\exp \left[ {-i\phi \left( {{\bf{R}},{\bf{r}},\tau } \right)} \right].
\end{eqnarray}
 Here, $\Delta$ is a real field, and ${\bf R}=(\mbox{\boldmath $r$}_a+ \mbox{\boldmath $r$}_b) /2$ and ${\bf r}=\mbox{\boldmath $r$}_a- \mbox{\boldmath $r$}_b$ are center of mass and relative coordinates, respectively. We assume a slow variation of the order parameter in space and time, and hence 
$\Delta \left( {{\bf{R}},{\bf{r}},\tau } \right) \cong \Delta \left( {\bf{r}} \right)$ and $
 \phi \left( {{\bf{R}},{\bf{r}},\tau } \right) \cong \phi \left( {{\bf{R}},\tau } \right)$ are satisfied. We  also introduce  auxiliary voltage field $V$ which coules to the charge operator. 
 Then, the partition function ${\cal Z}$ becomes 
\begin{widetext}
\begin{eqnarray}
 {\cal Z} = 
 \int {\cal D}\Delta _1 {\cal D}\Delta _2 {\cal D}\phi _1 {\cal D}\phi _2 {\cal D}V
 \exp \left[
     - \int_0^{\hbar \beta }  \frac{d\tau}{\hbar} \int  d \mbox{\boldmath $r$}_a d \mbox{\boldmath $r$}_b  
       \left\{ 
                 \frac{ \left| \Phi _1  \right|^2} 
                        {g_1 \left( {\mbox{\boldmath $r$}_a  -\mbox{\boldmath $r$}_b } \right)}   
             + \left( {1 \leftrightarrow 2} \right) 
            + \frac{CV^2}{2} 
                   \right\} 
                    + \mbox{Tr} \ln G^{ - 1} 
       \right] .
\end{eqnarray}
\end{widetext}
Here, $G$ is a $4 \times 4$ matrix Green's function
\begin{eqnarray}
 G^{ - 1} \left( {{\mbox{\boldmath $r$}}_a ,\tau _a ,{\mbox{\boldmath $r$}}_b ,\tau _b } \right) = \left( {\begin{array}{*{20}c}
   {\hat G_1^{ - 1} \left( {{\mbox{\boldmath $r$}}_a ,\tau _a ,{\mbox{\boldmath $r$}}_b ,\tau _b } \right)} & { - \hat t\left( {{\mbox{\boldmath $r$}}_a ,{\mbox{\boldmath $r$}}_b } \right)\delta \left( {\tau _a  - \tau _b } \right)}  \\
   { - \hat t^\dag  \left( {{\mbox{\boldmath $r$}}_a ,{\mbox{\boldmath $r$}}_b } \right)\delta \left( {\tau _a  - \tau _b } \right)} & {\hat G_2^{ - 1} \left( {{\mbox{\boldmath $r$}}_a ,\tau _a ,{\mbox{\boldmath $r$}}_b ,\tau _b } \right)}  \\
\end{array}} \right), 
\ \ \ 
\end{eqnarray}
where, 
\begin{eqnarray}
 \hat t\left( {{\mbox{\boldmath $r$}}_a ,{\mbox{\boldmath $r$}}_b } \right) =\left( {\begin{array}{*{20}c}
   {t\left( {{\mbox{\boldmath $r$}}_a ,{\mbox{\boldmath $r$}}_b } \right)e^{i\left( {\phi _1 ({\bf{R}},\tau ) - \phi _2 ({\bf{R}},\tau )} \right)/2} } & 0  \\
   0 & { - t^ *  \left( {{\mbox{\boldmath $r$}}_a ,{\mbox{\boldmath $r$}}_b } \right)e^{ - i\left( {\phi _1 ({\bf{R}},\tau ) - \phi _2 ({\bf{R}},\tau )} \right)/2} }  \\
\end{array}} \right),
\ \ \ 
\end{eqnarray}
\begin{widetext}
\begin{eqnarray}
\hat G_{1(2)}^{ - 1}  = \left[ { - \hbar \frac{\partial }{{\partial \tau }}   \hat\tau _0 + i\hbar \left( {v_{1(2)}  \cdot \nabla } \right)   \hat\tau _0 + \left\{ {\frac{{\hbar ^2 \nabla ^2 }}{{2m}} + \mu  - \frac{m}{2}v_{1(2)}^2  + i\left( {\frac{\hbar }{2}\frac{{\partial \phi _1 }}{{\partial \tau }} - eV_{1(2)} } \right)} \right\}\hat \tau _3  - \hat \Delta _{1(2)} } \right]\delta \left( {{\mbox{\boldmath $r$}}_a  - {\mbox{\boldmath $r$}}_b } \right)\delta \left( {\tau _a  - \tau _b } \right).
\label{eqn:green}
\end{eqnarray}
\end{widetext}

In the last equation, we introduced the Pauli matrix $\hat{\tau}_i$, the identity matrix $\hat{\tau}_0$, 
the pair potential
\begin{eqnarray}
 \hat \Delta _{1(2)} &=& 
  \left(
 \begin{array}{cc}
  0 & \Delta_{1(2)}({\bf{r}} ) e^{- i \phi_{1(2)}({\bf{R}},\tau )} \\
  \Delta_{1(2)}({\bf{r}} ) e^{ i \phi_{1(2)}({\bf{R}},\tau )} & 0
 \end{array}
 \right),
 \nonumber
\end{eqnarray}
the superfluid velocity $v_{1(2)}  =  - (\hbar / 2m) \nabla \phi _{1(2)}$, and 
$V_{1(2)}  =  + ( - )V/2$.
In the following, we assume that the phase varies slowly in space, namely $v_{1(2)} \approx 0$. 
The functional integrals over the modulus of the order parameter field $\Delta$ and the voltage field $V$  are performed by the saddle-point method, which lead  to the gap equation and the Josephson equation, respectively.\cite{Eckern,Ambegaokar,Zaikin} We also assume that the tunneling matrix element $t$ is finite only in the vicinity of the insulating barrier. 
Then, the partition function ${\cal Z}$ is reduced to a single functional integral over the phase difference $\phi=\phi_1-\phi_2$, 
\begin{eqnarray}
{\cal Z}  = \int {{\cal D}\phi \exp \left[ { - \frac{1}{\hbar }\int_0^{\hbar \beta } {\frac{C}{2}\left( {\frac{\hbar }{{2e}}\frac{{\partial \phi }}{{\partial \tau }}} \right)^2 d\tau }  + \mbox{Tr}\ln G_{}^{ - 1} } \right]}.
\end{eqnarray}

Next, we expand $\ln G^{-1}$ up to the second order in the tunneling matrix element $t$, i.e., 
\begin{eqnarray}
\mbox{Tr} \ln G_{}^{ - 1}  \approx \mbox{Tr} \ln \hat {\underline{g}}^{ - 1}  - \frac{1}{2} \mbox{Tr} \left( {\hat {\underline{g}} \thinspace  \hat{\underline{t}} \thinspace \hat {\underline{g}}} \thinspace \hat{\underline{t}} \right)
\end{eqnarray}
with  
\begin{eqnarray}
 \hat{\underline{g}}^{ - 1}  = \left( {\begin{array}{*{20}c}
   {\hat G_1^{ - 1} \left( {{\mbox{\boldmath $r$}}_a ,\tau _a ,{\mbox{\boldmath $r$}}_b ,\tau _b } \right)} & 0  \\
   0 & {\hat G_2^{ - 1} \left( {{\mbox{\boldmath $r$}}_a ,\tau _a ,{\mbox{\boldmath $r$}}_b ,\tau _b } \right)}  \\
\end{array}} \right), \\ 
 \hat{\underline{t}} = \left( {\begin{array}{*{20}c}
   0 & {\hat t\left( {{\mbox{\boldmath $r$}}_a ,{\mbox{\boldmath $r$}}_b } \right)\delta \left( {\tau _a  - \tau _b } \right)}  \\
   {\hat t^\dag  \left( {{\mbox{\boldmath $r$}}_a ,{\mbox{\boldmath $r$}}_b } \right)\delta \left( {\tau _a  - \tau _b } \right)} & 0  \\
\end{array}} \right).
\end{eqnarray}
After some calculations, we obtain 
\begin{eqnarray}
{\cal Z}
= 
\int 
{\cal D} \phi (\tau) 
\exp
\left(
  - \frac{{\cal S}_{\mathrm{eff}}[\phi]}{\hbar}
\right)
,
\end{eqnarray}
where the effective action ${\cal S}_{\mathrm{eff}}$ is given by~\cite{rf:Bruder95,rf:Barash95,Kawabata1} 
\begin{eqnarray}
{\cal S}_{\mathrm{eff}}[\phi]
&=&
\int_{0}^{\hbar \beta} d \tau 
\frac{C}{2}
\left(
   \frac{\hbar}{2 e} \frac{\partial \phi ( \tau) }{\partial \tau}
\right)^2
\nonumber\\
&-&
\int_{0}^{\hbar \beta} \int_{0}^{\hbar \beta} d \tau 
d \tau'
\left[
  \alpha (\tau - \tau') \cos \frac{\phi(\tau) - \phi (\tau') }{2}
  \right.
  \nonumber\\
&-&  \left.
  \beta (\tau - \tau') \cos \frac{\phi(\tau) + \phi (\tau') }{2}
\right]
.
\label{eqn:action}
\end{eqnarray}
The second term in eq. (\ref{eqn:action}) describes the dissipation due to the quasiparticle tunneling.
The third term describes the tunneling of Cooper pairs (the Josephson tunneling).
The only difference of the effective action between $d$-wave and $s$-wave junctions is the dependence of $\Delta$ on the relative coordinate ${\bf r}$, and is included in the two kernels.
The dissipation kernel $\alpha(\tau)$ and  the Josephson kernel $\beta(\tau)$ are given in terms of the diagonal and off-diagonal components of the Matsubara Green functions in Nambu space,  denoted by ${\cal G}$ and ${\cal F}$,
\begin{eqnarray}
  \alpha (\tau )  &=&  
  -\frac{2}{\hbar}
  \sum_{\mbox{\boldmath $k$},\mbox{\boldmath $k$}'}
  \left|
   t (\mbox{\boldmath $k$},\mbox{\boldmath $k$}')
  \right|^2
  {\cal G}_1 \left( \mbox{\boldmath $k$},\tau \right)
 {\cal G}_2 \left( \mbox{\boldmath $k$}',-\tau \right)
 ,
 \\
  \beta (\tau ) &=& 
  -\frac{2}{\hbar}
  \sum_{\mbox{\boldmath $k$},\mbox{\boldmath $k$}'}
  \left|
   t (\mbox{\boldmath $k$},\mbox{\boldmath $k$}')
  \right|^2
  {\cal F}_1 \left( \mbox{\boldmath $k$},\tau \right)
 {\cal F}_2^\dagger \left( \mbox{\boldmath $k$}',-\tau \right)
 .\label{beta}
\end{eqnarray}
The Green functions are given by 
\begin{eqnarray}
 {\cal G} \left( \mbox{\boldmath $k$},\omega_n \right)
 &=&- 
  \frac{\hbar \left( i \hbar \omega_n + \xi_{\mbox{\boldmath $k$}} \right)}
  {(\hbar \omega_n)^2 + \xi_{\mbox{\boldmath $k$}}^2 +  \Delta (\mbox{\boldmath $k$})^2} 
  ,
  \\
 {\cal F} \left( \mbox{\boldmath $k$},\omega_n \right)
 &=&
  \frac{\hbar \Delta (\mbox{\boldmath $k$})}
  {(\hbar \omega_n)^2 + \xi_{\mbox{\boldmath $k$}}^2 +  \Delta (\mbox{\boldmath $k$})^2}
,  \label{f}
  \end{eqnarray}
where $\xi_{\mbox{\boldmath $k$}}=\hbar^2 \mbox{\boldmath $k$}^2/2 m -\mu$ and $\hbar \omega_n= (2n+1) \pi /\beta$ is the fermionic Matsubara frequency ($n$ is an integer).
Information about the anisotropy of  the order parameter is included in $\Delta (\mbox{\boldmath $k$})$.
In the case of the cuprate high-$T_c$ superconductors (the $d_{x^2-y^2}$ OPS), $\Delta(\mbox{\boldmath $k$})$ is given by 
$  
\Delta (\mbox{\boldmath $k$})
  = 
  \Delta_0 \cos 2 \theta
$
. Here, $\theta$ is an angle between $\mbox{\boldmath $k$}$ and  $a$-axis. 

\subsection{Effective action for current-biased $c$-axis twist junctions}
Let us turn to the calculation of the effective action for the $c$-axis twist Josephson junctions (see Fig. 1).
We define $\gamma$ as the twist angle about the $c$-axis ($0 \le \gamma < \pi/4$).~\cite{Klemm} 
Such junctions can be fabricated by using the single crystal whisker of Bi2212.~\cite{Takano1,Takano2,Badica,Latyshev}
Takano {\it et al.} measured the $\gamma$ dependence of the $c$-axis Josephson critical current and showed a clear evidence of the $d_{x^2-y^2}$ OPS of Bi2212.~\cite{Takano1,Takano2,Maki}

In what follows, we assume that the tunneling between the two superconductors is described in terms of the coherent tunneling, i.e.,
\begin{eqnarray}
\left| t(\mbox{\boldmath $k$},\mbox{\boldmath $k$}')\right|^2 
 = 
 \left| t \right|^2 \delta_{\mbox{\boldmath $k$}_\parallel,\mbox{\boldmath $k$}'_\parallel},
 \label{coherent-tunnel}
 \end{eqnarray}
where $\mbox{\boldmath $k$}_\parallel$ is the momentum parallel to the $ab$-plane.
For simplicity, we also assume that each superconductor consists of single  CuO${}_2$ layer, $\Delta_1 (\mbox{\boldmath $k$})=\Delta_0 \cos 2 \theta$, and $\Delta_2 (\mbox{\boldmath $k$})=\Delta_0 \cos 2 \left( \theta + \gamma \right)$ (see Fig. \ref{f1}).
Moreover, we consider the low temperature limit ($k_B T \ll \Delta_0$).
In the case where the  Josephson junction is biased by an externally applied current $I_{\mathrm{ext}}$, we have to add an additional potential contribution linear in $\phi$.~\cite{Ambegaokar,Eckern}
At this level of approximation, the effective action ${\cal S}_{\mathrm{eff}}$ of the current biased $c$-axis twist Josephson junction has the form 
\begin{eqnarray}
{\cal S}_{\mathrm{eff}}[\phi]
&= &
\int_{0}^{\hbar \beta} d \tau 
\left[
   \frac{M}{2} 
   \left(
   \frac{\partial \phi ( \tau) }{\partial \tau}
   \right)^2
   + 
   U(\phi)
\right]
+
{\cal S}_\mathrm{diss}[\phi]
,
\nonumber\\
\\
{\cal S}_\mathrm{diss}[\phi]
&= &
-
\int_{0}^{\hbar \beta} \int_{0}^{\hbar \beta} d \tau 
 d \tau'
  \alpha (\tau - \tau') \cos \frac{\phi(\tau) - \phi (\tau') }{2}
  , \label{action}
\nonumber\\
\end{eqnarray}
where $M=C(\hbar/2e)^2$ is the mass and $ U(\phi) $ is the tilted washboard potential
\begin{eqnarray}
U(\phi)=  - E_J  \left[ \cos \phi( \tau)  + \eta \phi(\tau) \right].
\end{eqnarray}
In this equation, $\eta$ is given by $\eta = I_\mathrm{ext}/ I_C  $, and $E_J=\left(  \hbar/2 e \right) I_C$ is the Josephson coupling energy, where
\begin{eqnarray}
I_C = - \frac{2e}{\hbar} \int_0^{\hbar \beta} d \tau \beta (\tau) 
\label{ic}
\end{eqnarray}
is the Josephson critical current.
In the derivation of the Josephson term, we adopted the local approximation.\cite{rf:MQT1,Joglekar} Details of this
approximation and its justification are given in Appendix ~\ref{sec:app_beta}.

\begin{figure}[b]
\begin{center}
\scalebox{0.4}{
\includegraphics[width=18.0cm,clip]{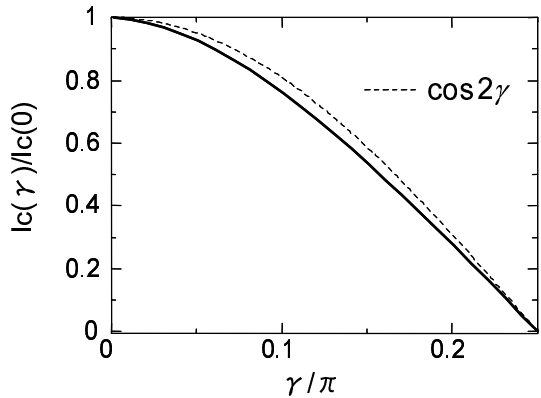}}
\end{center}
\caption{ The twist angle $\gamma$ dependence of the critical current $I_C$. 
Broken line is proportional to $\cos2\gamma$. 
} \label{f2}
\end{figure}

By substituting Eq. (\ref{beta}) into Eq. (\ref{ic}), we can obtain the $\gamma$ dependence of $I_C$, i.e., 
\begin{eqnarray}
 I_C\left( \gamma  \right)  
 & =& \frac{{2e}}{\hbar }\left| t \right|^2 N_{0}^2 \Delta_0 \int_0^{2\pi } { \frac{d\theta}{{2\pi }} \cos 2\theta \cos } 2\left( {\theta  + \gamma } \right)  \nonumber \\
 & \times& \int_0^\infty  {dy\frac{1}{{\sqrt {y^2  + \cos ^2 2\theta } }}\frac{1}{{\sqrt {y^2  + \cos ^2 2\left( {\theta  + \gamma } \right)} }}} 
 ,
  \nonumber \\
\end{eqnarray}
where $N_0$ is the density of states at the Fermi energy. 
Especially at $\gamma=0$, we get a well-known result~\cite{KashiwayaTanaka}
\begin{eqnarray}
I_C (0)&=&\frac{2 e}{\hbar}|t|^2 N_0^2 \Delta_0.
\label{ic0}
\end{eqnarray}
Figure \ref{f2} shows the twist angle $\gamma$ dependence of the critical current $I_C$. 
It is approximately proportional to $\cos2\gamma$.
Note that for $\gamma > \pi/4$, the sign of the current changes. 
Thus, in this case, the $\pi$-junction is formed.
These behaviors are attributed to the $d$-wave OPS of high-$T_c$ superconductors.

Next, we will calculate the dissipation kernel $\alpha(\tau) $ for the $c$-axis twist junctions.
In the case of the $c$-axis junction with $\gamma=0$ (e.g., single intrinsic junctions),~\cite{You1,You2}  the node-to-node quasiparticle tunneling is always possible.
Then, the explicit and asymptotic form of $\alpha(\tau) $ at low temperature are obtained as
\begin{eqnarray}
\alpha(\tau) 
&=&
\frac{{\left| t \right|^2 N_{0}^2 \Delta_0 ^2 }}{{2\pi ^2 \hbar }}
 \int_0^{2\pi }
 d\theta  \cos^2 2\theta
 K_1 \left( {\frac{{\left| \tau  \right|}}{\hbar }\Delta_0 \left| {\cos 2\theta } \right|} \right)^2
\label{superohmic}
\nonumber\\
\\
&\approx&
\frac{{3\hbar ^2  }}{{16\pi \Delta _0^{} }} \frac{R_Q}{R}  \frac{1}{{\left| \tau  \right|^3 }} \quad  \mathrm{for}  \quad \frac{\Delta_0 |\tau| }{\hbar} \gg1,
\label{superohmic2}
\end{eqnarray}
where $K_1$ is the modified Bessel function, 
 $R_Q  = h/4e^2$ is the resistance quantum and 
\begin{eqnarray}
\frac{1}{R} = \frac{{2 e^2 }}{\hbar }\left| t \right|^2 N_0^2.
\end{eqnarray}
is the inverse of the normal state resistance $R$.
Equation (\ref{superohmic2}) gives the superohmic dissipation which agrees with the previous results~\cite{rf:Bruder95,rf:Barash95,Kawabata1}  based on the coherent tunneling approximation (Eq. (\ref{coherent-tunnel})).
The present model shows stronger dissipation effect than the incoherent tunneling model ($\alpha(\tau) \sim 1/ \tau^4$)~\cite{Joglekar} and a more realistic tunneling model based on the first-principle band-calculation\cite{Anderson} ($\alpha(\tau) \sim 1/ |\tau|^5$).~\cite{Tanaka} 
Despite this fact, the quasiparticle dissipation effect on MQT in the present coherent tunneling model is still quite weak as will be shown in Sec. III. 
See Appendix B for the derivation of these asymptotic forms of the kernel $\alpha(\tau)$.

In the case of nonzero $\gamma$, we get
\begin{eqnarray}
 \alpha \left( \tau  \right) 
 &= &
 \frac{{\left| t \right|^2 N_{0}^2 \Delta_0 ^2 }}{{2\pi ^2 \hbar }}
 \int_0^{2\pi }
 {d\theta \left| {\cos 2\theta \cos 2\left( {\theta  + \gamma } \right)} \right|}
 \nonumber \\
&\times&
 K_1 \left( {\frac{{\left| \tau  \right|}}{\hbar }\Delta_0 \left| {\cos 2\theta } \right|} \right)
 K_1 \left( {\frac{{\left| \tau  \right|}}{\hbar }\Delta_0 \left| {\cos 2\left( {\theta  + \gamma } \right)} \right|} \right) .
\nonumber
\\
\end{eqnarray}
The asymptotic behavior of $\alpha(\tau)$ for nonzero $\gamma$ is numerically estimated as  an exponential function, i.e.,
\begin{eqnarray}
\alpha(\tau)
\sim 
  \exp \left[ - \sin \left (2 \gamma \right)  \frac{\Delta_0 |\tau| }{\hbar} \right], \label{kernel}
\end{eqnarray}
for $\Delta_0 |\tau| /\hbar \gg1$.
This exponential  behavior is ascribed to the suppression of the node-to-node quasiparticle tunneling by the finite value of $\gamma$~\cite{rf:Bruder95,Kawabata1} and is similar to that of the conventional $s$-wave junctions with the constant order parameter $\Delta$:
$
\alpha (\tau) \sim
 \exp \left( - 2 \Delta |\tau| /\hbar \right)
.$~\cite{Ambegaokar,Eckern}
For nonzero $\gamma$, if the phase $\phi(\tau)$ varies slowly with time on the scale given by $\hbar/\Delta_0$, we can expand $\phi(\tau)-\phi(\tau')$ in Eq.(\ref{action}) with respect to $\tau-\tau'$, which results in 
\begin{eqnarray}
{\cal S}_\mathrm{diss} [\phi]
\approx
\frac{\delta M (\gamma)}{2} 
\int_0^{\hbar \beta} d \tau
\left(  \frac{\partial \phi (\tau) }{\partial \tau} \right)^2
,
\end{eqnarray}
where 
\begin{widetext}
\begin{eqnarray}
 \delta M\left( \gamma  \right) &=& \frac{1}{2}\int_0^\infty  {d\tau \tau ^2 } \alpha \left( \tau  \right) 
  = \frac{{\hbar ^2 \left| t \right|^2 N_{0}^2 }}{\Delta_0 } \int_0^{2\pi } {\frac{d\theta}{2 \pi}  \frac{{\left[ {\cos ^2 2\theta  + \cos ^2 2\left( {\theta  + \gamma } \right)} \right] E\left( k \right) - 2\left| {\cos 2\theta \cos 2\left( {\theta  + \gamma } \right)} \right|K\left( k \right)}}{{\left[ {\left| {\cos 2\theta } \right| + \left| {\cos 2\left( {\theta  + \gamma } \right)} \right|} \right]\left[ {\left| {\cos 2\theta } \right| - \left| {\cos 2\left( {\theta  + \gamma } \right)} \right|} \right]^2 }}} \label{mass}
  \nonumber\\
\end{eqnarray}
\end{widetext}
for $\gamma \ne 0$ 
with $k= [\left| {\cos 2\theta } \right| - \left| {\cos 2\left( {\theta  + \gamma } \right)} \right|]/[{\left| {\cos 2\theta } \right| + \left| {\cos 2\left( {\theta  + \gamma } \right)} \right|}]$. 
Here, $E(k)$ and $K(k)$ are complete elliptic integral of the second and the first kinds, respectively. 
Hence, under the above condition, the dissipation action ${\cal S}_\mathrm{diss}$ acts as a kinetic term so that the effect of the quasiparticle dissipation results in an increase  of the mass, i.e., $M \to M + \delta M$ (the mass renormalization).
This result indicates that the influence of the quasiparticle dissipation for nonzero $\gamma$ is quite weaker than that for $\gamma=0$, and hence it also gives negligible contribution to MQT as will be shown in the next section.
 It should be remarked that the increase of the mass $\delta M$ is inversely proportional to $\Delta_0$.

\subsection{MQT rate}

We will calculate the $\gamma$ dependence of the MQT rate at low temperatures. 
Note that, in the previous studies of MQT in $d$-wave $c$-axis junctions,~\cite{Kawabata1,Kawabata3,Kawabata4} only the limiting cases of $\gamma$, i.e., $\gamma=0$ and $\pi/8$, have been considered. 
The MQT rate is defined by the formula~\cite{rf:MQT1,rf:MQT2,rf:Coleman}
\begin{eqnarray}
\Gamma= 
\lim_{\beta \to \infty} \frac{2}{\beta} \mbox{ Im}\ln {\cal Z}
.
\end{eqnarray}
In order to determine $\Gamma$, we employ the bounce approximation.~\cite{rf:MQT1,rf:MQT2,rf:Coleman} 
When the barrier is low enough for the MQT to occur but still so high that the bounce approximation can be valid, $\Gamma$ is given by
\begin{eqnarray}
\Gamma 
\approx
A
\exp \left( - \frac{S_B}{\hbar} \right)
,
\end{eqnarray}
where $S_B= {\cal S}_{\mathrm{eff}}[\phi_B]$ is the bounce exponent, which is the value of the the action ${\cal S}_{\mathrm{eff}}$ evaluated along the bounce trajectory $\phi_B(\tau)$.
Assuming  that $I_{\mathrm{ext}}$ is close to $ I_C$, we can approximate the washboard potential $U(\phi)$ as a quadratic-puls-cubic one. 
Then, we obtain for $\gamma  = 0$\cite{Kawabata1,Kawabata3,Kawabata4}
\begin{eqnarray}
\thinspace
\Gamma(0,\eta)
&\approx&
\Gamma_0(0,\eta)
 \exp \left[ 
 -B(0,\eta)
 - c_0
  \frac{ \hbar I_C(0)}{\Delta_0^2}
  \sqrt{ \frac{\hbar}{2 e} \frac{I_C(0)}{C}}
   \right.
\nonumber\\
&&
 \times
\left.
 \left(
  1 -
  \eta^{2}
\right)^{5/4}
  \right], 
  \label{MQT0}
\end{eqnarray}
 and 
for $\gamma \neq 0 $
\begin{eqnarray}
\Gamma(\gamma,\eta)
&\approx&
\Gamma_0(\gamma,\eta)
 \exp 
 \left[
    - B\left(\gamma,\eta \right)
\right]
.
  \label{MQT1}
\end{eqnarray}
In Eqs. (\ref{MQT0}) and (\ref{MQT1}), $\eta= I_\mathrm{ext} / I_C (\gamma)$,  $c_0 = (27 \pi/8) $ $\int_0^\infty d x [x^4/\sinh^2(\pi x) ] \ln (1+ x^{-2}) \approx 0.14$, 
\begin{eqnarray}
B(\gamma,\eta)
&=&
\frac{6}{5e} \sqrt{ \frac{\hbar}{2 e} I_C(\gamma) C}
\frac{\delta M  (\gamma)}{M}
	\left(
1-
            \eta^{2}
	\right)^{5/4} ,
\end{eqnarray}
and $\Gamma_0(\gamma,\eta)$ is the decay rate without the quasiparticle dissipation, i.e.,
\begin{eqnarray}
 \Gamma _0(\gamma,\eta)  = 12\omega _p (\gamma,\eta) \sqrt {\frac{{3U_0 (\gamma,\eta) }}{{2\pi \hbar \omega _p (\gamma,\eta) }}} \exp \left[ { - \frac{{36U_0  (\gamma,\eta)}}{{5\hbar \omega _p (\gamma,\eta) }}} \right], 
 \nonumber\\
 \end{eqnarray}
where barrier height $U_0$ and the Josephson plasma frequency $ \omega _p$ are given by
\begin{eqnarray}
 U_0  (\gamma,\eta) &=& \frac{{\hbar I_C (\gamma) }}{{3e}}\left( {1 - \eta ^2 } \right)^{3/2},  \\ 
 \omega _p  (\gamma,\eta)& = &\sqrt {\frac{{\hbar I_C (\gamma) }}{{2eM}}} \left( {1 - \eta ^2 } \right)^{1/4}  .
\end{eqnarray}
For $\gamma=0$, $\delta M(0)$ is the mass increment due to the high-frequency components ($\omega > \omega_p$) of the dissipation kernel (Eq. (\ref{superohmic})) and is given by
\begin{eqnarray}
\delta M(0)=\frac{\hbar^2 N_0^2 |t|^2}{\pi^2 \Delta_0} \! \int_0^1\!  d x \frac{x^2}{\sqrt{1-x^2}} \int _0^{\frac{\Delta_0}{\hbar \omega_p (0,\eta)}} \!  d s s^2 K_1(s x)^2
.
\nonumber\\
\end{eqnarray}
The second term in the exponent of Eq. (\ref{MQT0}) results from the low-frequency components ($\omega < \omega_p$) of the dissipation kernel $\alpha(\tau)$ (Eq. (\ref{superohmic})).

\subsection{Switching current distribution and crossover temperature}

The switching current distribution $P(\eta)$ is related to the MQT rate $\Gamma(\gamma, \eta)$ as~\cite{rf:Voss81,Garg} 
\begin{eqnarray}
P(\eta)=\frac{1}{v}
\Gamma(\gamma, \eta) \exp
\left[
- \frac{1}{v}
\int_0^{\eta} \Gamma(\gamma,\eta') d \eta'
\right]
,
\end{eqnarray}
where $v \equiv \left| d \eta / d t \right| $ is the sweep rate of the external bias current.
The mean value $\langle \eta \rangle$ and the square mean value $\langle \eta^2 \rangle$ of the switching current are respectively expressed by
\begin{eqnarray}
\langle \eta \rangle &\equiv& \int_0^1 d \eta' P(\eta') \eta',
\\
\langle \eta^2 \rangle &\equiv& \int_0^1 d \eta' P(\eta') {\eta'}^2.
\end{eqnarray}
Then, the standard deviation $\sigma_n$ of the switching current distribution $P(\eta)$ is defined by
\begin{eqnarray}
\sigma_n  = I_{C} \sqrt {\left\langle {\eta ^2 } \right\rangle  - \left\langle \eta  \right\rangle ^2 } .
\end{eqnarray}
In actual MQT experiments, the temperature $T$ dependence of $\sigma_n$ is measured.
At high temperature, the TA decay dominates the escape process.
Then, the escape rate $\Gamma$ is given by the Kramers formula~\cite{rf:MQT1,rf:MQT2,Kramers}
\begin{eqnarray}
\Gamma (\gamma,\eta)=\frac{\omega_p  (\gamma,\eta)}{2 \pi }\exp \left[ - \frac{U_0  (\gamma,\eta)}{ k_B T }\right].
\end{eqnarray}
Below  $T^*$, the escape process is dominated by MQT. 
The crossover temperature $T^*$ is determined from an intersection point of  $\sigma_n$ between $T$-independent MQT and $T$-dependent TA process.
In the low-dissipation (underdamping) cases, $T^*$ is approximately given by~\cite{Grabert,Hanggi,Kato}
\begin{eqnarray}
T^* (\gamma)
=\frac{  \hbar \omega_p (\gamma,\eta=\langle \eta \rangle)}{2 \pi k_B}
  \label{eqn:Tco}.
\end{eqnarray}

\section{Results}

In this section, we will numerically calculate the MQT rate $\Gamma$ and  the crossover temperature $T^*$ for the $c$-axis twist junctions.
In the following we choose $v=42.4$mA/s/$I_C(0)$, $I_C(0)=48.54\mu$A, $C=76.26$fF and $\Delta_0=30$meV unless explicitly mentioned. 
These parameters were used in a MQT experiment of Bi2212 intrinsic junctions.~\cite{Inomata}

\begin{figure}[tb]
\begin{center}
\scalebox{0.4}{
\includegraphics[width=18.0cm,clip]{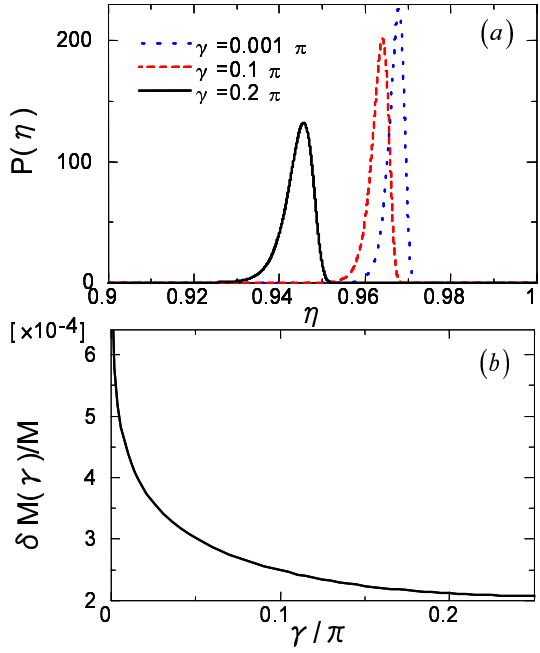}}
\end{center}
\caption{ (color online) The twist angle $\gamma$ dependences of  (a) the switching current distribution $P(\eta) $ and (b) the mass increment $\delta M$  due to the quasiparticle dissipation.
} \label{f3}
\end{figure}

Figure \ref{f3} (a) and Figure \ref{f3} (b) exhibit the $\gamma$ dependences of the switching current distribution $P(\eta)$ and the mass increment $\delta M$ due to the quasiparticle dissipation, respectively. As $\gamma$ increases, $P(\eta)$ becomes smeared, and the peak position shifts to the smaller values of $\eta$. 
This is because $I_C$ decreases with increasing $\gamma$, which makes the switching process more likely to occur even for  small $\eta$. 
As shown in Fig. \ref{f3} (b), the ratio $\delta M/M$ is of the order of $10^{-4} \ll 1$  for all $\gamma$.
Therefore, the mass renormalization effect in twist junctions is quite weak, especially for large $\gamma$. 
This $\gamma$ dependence can be explained as follows.
As  $\gamma$ increases, the node-to-node quasiparticle tunneling becomes more suppressed.
 Therefore, the effect of the quasiparticle dissipation, namely the mass increment $\delta M$, is decreasing with increasing  $\gamma$. 

\begin{figure}[t]
\begin{center}
\scalebox{0.4}{
\includegraphics[width=20.0cm,clip]{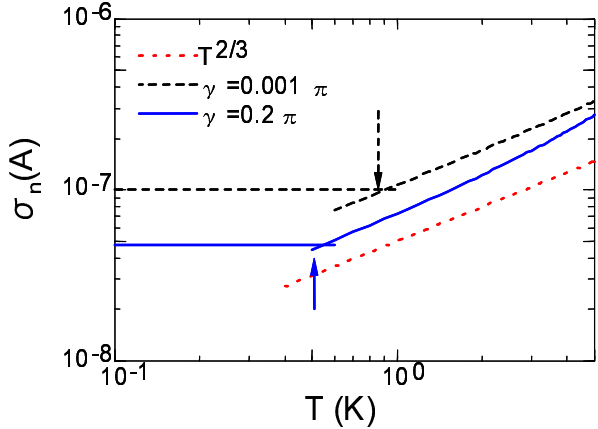}}
\end{center}
\caption{ (color online) The standard deviation $\sigma_n$ of the switching current distribution $P(\eta)$ for various $\gamma$. 
Doted line is proportional to $T^{2/3}$. Arrows point the crossover temperature $T^*$ calculated from the approximate formula (\ref{eqn:Tco}). 
} \label{f4}
\end{figure}

\begin{figure}[h]
\begin{center}
\scalebox{0.4}{
\includegraphics[width=16.0cm,clip]{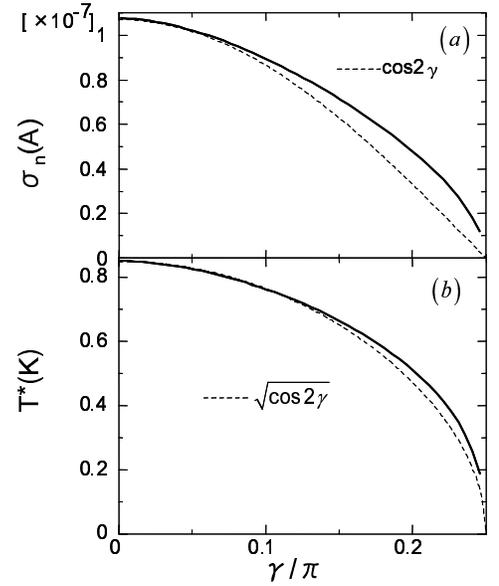}}
\end{center}
\caption{ The twist angle $\gamma$  dependences of (a) the standard deviation $\sigma_n$ of the switching current distribution $P(\eta)$ and  (b) the crossover temperature $T^*$. Broken lines are proportional to $\cos2\gamma$ and $\sqrt{\cos2\gamma}$ in (a) and (b), respectively. 
} \label{f5}
\end{figure}

In Fig. \ref{f4}, the standard deviation $\sigma_n$ of the switching current distribution $P(\eta)$ is depicted for various $\gamma$.  
The crossover temperature $T^*$ decreases with increasing $\gamma$.
 In the TA regime, the switching current is proportional to $T^{2/3}$.~\cite{Garg,Kramers}
In Fig. \ref{f5}, we show $\gamma$ dependences of (a) $\sigma_n$ and (b) $T^*$. 
They decrease monotonically with the increase of $\gamma$. Their $\gamma$ dependences are approximately given by $\cos2\gamma$ and $\sqrt{\cos2\gamma}$, respectively.
The $\gamma$ dependence of $T^*$ can be explained as follows. 
Due to the $d$-wave OPS,  $I_C (\gamma)$ is nearly proportional to $\cos2\gamma$ as shown in Fig. \ref{f2}.
Thus, we obtain 
\begin{eqnarray}
T^* \propto  \omega_p \propto \sqrt{I_C (\gamma)}  \propto \sqrt{\cos2\gamma} 
\end{eqnarray}
if we neglect the $\gamma$ dependence of $\langle \eta \rangle$. 
Hence, $T^*$ is almost proportional to $\sqrt{\cos2\gamma}$.

\begin{figure}[t]
\begin{center}
\scalebox{0.4}{
\includegraphics[width=18.0cm,clip]{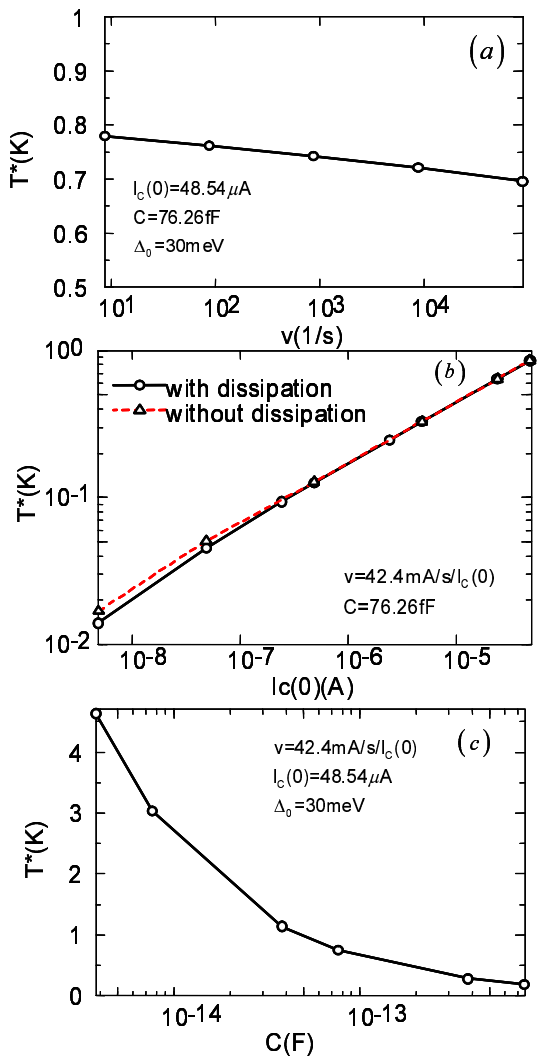}}
\end{center}
\caption{(color online) The dependences of the crossover temperature $T^*$ on (a) the sweep rate $v$, (b) the critical current $I_C(0)$ which is proportional to $\Delta_0$(we set $I_C(0)=48.54\mu$A at $\Delta_0=30$meV), and (c) the capacitance $C$ for $c$-axis junction with $\gamma=0$.
} \label{f6}
\end{figure}

For applications to, e.g., $d$-wave phase qubit,~\cite{Kawabata7} higher $T^*$ is desirable for high temperature qubit operation. 
Thus, we will study the dependences of $T^*$ on other parameters ($v$, $I_C$ and $C$) by changing these parameters, and clarify the condition for realizing high $T^*$ in the case of $\gamma=0$. 

Figure \ref{f6} (a) shows the sweep rate $v$ dependence of $T^*$. 
Note that $v$ has to be much smaller than the plasma frequency $\omega_p$ in order to apply our theory. 
The crossover temperature $T^*$  is an decreasing function of $v$.
 However, it quite weakly depends on $v$. 
 More sensitive parameters are the critical current $I_C(0)$ and capacitance $C$. 
 Figure \ref{f6} (b) displays the $I_C(0)$ dependence of $T^*$, where  $I_C(0) \propto \Delta_0$ (see Eq. (\ref{ic0})).
  As $I_C$ increases,  $T^*$ increases. 
On the other hand,  $T^*$ decreases with increasing $C$ as shown in Fig. \ref{f6} (c). 
The obtained parameter dependences of $T^*$ are consistent with the simple formula, $T^* \propto \sqrt{I_C/C}$.
In Fig. \ref{f6} (b), we also show the result without the quasiparticle dissipation effect. 
Since the ratio $\delta M/M$ is inversely proportional to $\Delta_0$, the influence of the quasiparticle dissipation becomes stronger for smaller $I_C(0)$. 
However, the effect of dissipation is still weak even for $I_C(0) \sim$ 1 nA, which is consistent with the recent experiment with a low $J_c$  Bi2212 
surface intrinsic Josephson junction.\cite{Li}
 
 Note that, for large $I_C$, the Josephson penetration depth $\lambda_J$ of the junction becomes small.
In the case of $I_C(0)=10^{-4}$ A, $\lambda_J$ is typically of the order of 1 $\mu$m.
In order to apply our MQT theory, the size of the junction should be smaller than $\lambda_J$.
It should be also noted that $T^*$ can approximately expressed by 
\begin{eqnarray}
T^*  \propto \sqrt {\frac{\Delta_0 }{{\varepsilon _r \rho }}} 
\end{eqnarray}
where $\varepsilon _r$ and $\rho$ are the relative permittivity and the normal-state resistivity of the junctions, respectively. 
Therefore, $T^*$ is almost independent of the size of the junction. 
Thus, in order to obtain high $T^*$, large magnitude of the gap $\Delta_0$ and small magnitudes of the relative permittivity $\varepsilon _r$ and the resistivity $\rho$ are desirable.

Finally, let us comment on the effect of the quasiparticle dissipation on MQT. 
As was mentioned above, this effect is quite weak even at $\gamma=0$, in which the node-to-node quasiparticle tunneling is possible. 
In fact, we have confirmed that even when the dissipation term is neglected, the obtained results are almost the same. 
Note that, in order to see the quasiparticle dissipation effect on MQT more clearly, we have to choose at least three orders smaller magnitude of the gap $\Delta_0$ as seen from Eq. (\ref{mass}) and Fig. \ref{f3} (b).

\section{Application to qubits}

In this section, we briefly discuss the advantage of $c$-axis twist junctions for the qubit application.

As was shown in the previous section, the quasiparticle dissipation on MQT is negligibly small.
This result strongly indicates the high potentiality of $c$-axis twist junctions for $d$-wave qubit.~\cite{Kawabata7,Ioffe,Blais,Blatter,Zagoskin,Tafuriqubit,Fominov,Amin04,Kolesnichenko,Amin05,Rotoli}

Moreover, in the case of the twist junction with $\gamma > \pi/4$, the $\pi$-junction is formed (see Sec. II.B).
A superconducting ring with the $\pi$-junction exhibits a spontaneous current without an external magnetic field and the corresponding magnetic flux is half a flux quantum $\Phi_0=\hbar/2e$ in the ground state.~\cite{pi}
Thus, the high-$T_c$ superconductor ring~\cite{Kim,Sandberg} including the $c$-axis twist junction with $\gamma > \pi/4$ becomes a quiet qubit~\cite{Ioffe,Blatter,Kawabata8-1,Kawabata8-2} that can be efficiently decoupled from the fluctuation of the external field.

In order to realize the $d$-wave quantum computer with large number of qubits, we have to tune the circuit parameter ($\omega_p$) of each qubit independently.
In actual $d$-wave junctions, however, precise control of $\omega_p$, e.g., by changing the thickness of the insulating barrier or by the oxygen doping, is very difficult.
Our results in Sec. III clearly show that we can artificially and precisely control the qubit parameter $\omega_p \propto T^*$ only by varying $\gamma$ of $c$-axis twist junctions.
Therefore, high-controllability of qubit parameters is another advantage of $c$-axis twist junctions for realizing scalable quantum computers.

\section{Summary}

In this paper, we have investigated the influence of the $d$-wave OPS on MQT in $c$-axis twist Josephson junctions by use of  the functional integral method and the bounce approximation. 
Due to the $d$-wave OPS, the twist angle $\gamma$ dependences of standard deviation $\sigma_n$ of the switching current distribution and the crossover temperature $T^*$ are approximately given by $\cos2\gamma$ and $\sqrt{\cos2\gamma}$, respectively.
Therefore, MQT in $c$-axis twist junction becomes a useful experimental tool for testing OPS of high-$T_c$ superconductors at low temperature regimes.
Moreover, the influence of the quasiparticle dissipation is found to be very weak.
This result indicates the high potential of $c$-axis twist junctions for qubit applications.

Throughout this paper, we have considered MQT in the single junctions where the quasiparticle dissipation effect is found to be negligibly small.
On the other hand,  in multiple stack intrinsic junctions, the long-range capacitive coupling between junctions~\cite{Koyama,Machida3} gives rise to the $O(N^2)$ enhancement of the MQT rate,~\cite{Jin,Machida1,Machida2,Fistul1,Fistul2,Nori} where $N$ is the number of the stack.
Therefore, the effect of the long-range capacitive coupling on the quasiparticle dissipation in multiple stack junctions will be an interesting subject of future studies.

\acknowledgments

We acknowledge valuable discussions with Y. Asano, A. A. Golubov, and S. Kashiwaya. 
S. K. would like to thank T. Claeson, P. Delsing, K. Inomata,  X. Y. Jin, H. Kashiwaya, T. Koyama, M. Machida, T. Matsumoto, P. M\"uller, F. Nori, S. Savel'ev, H. Shibata, V. S. Shumeiko, A. Y. Smirnov, F. Tafuri, Y. Takano, A. Tanaka, A. V. Ustinov, L. X. You, D. Winkler and A. M. Zagoskin for useful discussions.
T. Y. acknowledges support by the JSPS. 
This work was supported by a Grant-in-Aid for Scientific Research from the Ministry of Education, Science,
Sports and Culture of Japan (grant No. 17071007 and 17710081).
 This work was also supported by NAREGI Nanoscience Project, the Ministry of Education, Culture,
Sports, Science and Technology, Japan, the Core Research for Evolutional
Science and Technology (CREST) of the Japan Science and Technology
Corporation (JST) and a Grant-in-Aid for the 21st Century COE "Frontiers of
Computational Science". The computational aspect of this work has been
performed at the Research Center for Computational Science, Okazaki National
Research Institutes and the facilities of the Supercomputer Center,
Institute for Solid State Physics, University of Tokyo.

\appendix

\section{Nonlocality of the $\beta$ term}
\label{sec:app_beta}
The effective action in eq.~(\ref{eqn:action}) includes two terms, which are 
nonlocal in the imaginary time. In this paper, we have studied dissipation effects
caused by the second term (the $\alpha$ term) of eq.~(\ref{eqn:action}), while the third term (the $\beta$ term) has been treated by the local approximation: 
The Josephson coupling is calculated by assuming that the local component at 
$\tau \sim \tau'$ is dominant in the double integral of the $\beta$ term.
In this approximation, we can separate the integral by changing variables as $T=(\tau+\tau')/2$ and 
$t = \tau - \tau'$. Then, the Josephson energy is defined as
\begin{equation}
 E_J \equiv -\int_0^{\hbar \beta} dt \beta(t).
\end{equation}
As a result, the Josephson term in the effective action takes the usual form
\begin{equation}
- E_J \int_0^{\hbar \beta} dT \cos \phi(T).
\label{eq:act0}
\end{equation}
This approximation is justified in the $s$-wave superconducting junction
where the $\beta$-term has an exponential form ($\beta(\tau)\sim e^{- 2\Delta_0 |\tau|/\hbar}$).
In the high-$T_c$ junctions, however, the $\beta$ term shows a power-law decay,
and the local approximation is not justified in general. For example, it has been
discussed in ref.~\onlinecite{Khveshchenko1} that the ohmic
power-law decay of the $\beta$ term may affect the phase transition of the Josephson
junctions made of $d$-wave superconductors. In this appendix, 
we study the nonlocal effect of the $\beta$-term, and show that
the local approximation is actually justified in the calculation of the MQT rate.

Let us start with the effective action within the local approximation, which is given as
\begin{eqnarray}
S &=& \frac{\hbar^2}{4E_C} \int_0^{\hbar \beta} \left( \frac{\partial \phi(\tau)}{\partial \tau}
\right)^2 d\tau  \nonumber \\
&-& E_J \int_0^{\hbar \beta} d \tau (\cos \phi(\tau) + \eta \phi(\tau)).
\end{eqnarray}
Here, we neglected the dissipative term.
After changing the phase variable as $\psi = \phi - \pi/2 + 
\sqrt{2(1-\eta)}$, the action is written as
\begin{eqnarray}
S&=&\frac{\hbar^2}{4 E_C} \int_0^{\beta\hbar} d\tau \left(\frac{d\psi}{d\tau}\right)^2
\nonumber \\
& & \hspace{-8mm} + E_J \int_0^{\beta \hbar} d\tau \left[ - \frac16 \psi^3 + \frac12
\sqrt{2(1-\eta)} \psi^2 \right],
\end{eqnarray}
where $E_c = (2e)^2/2C$.
Here, the potential term is approximated by a cubic polynomial. We further rescale the
variables as $\tilde{\tau} = \omega_0 \tau$ and $\psi = \sqrt{2(1-\eta)} \tilde{\psi}$,
where $\hbar \omega_0 = \sqrt{2E_C E_J} (2(1-\eta))^{1/4}$. Then, 
the action is obtained as
\begin{eqnarray}
\frac{S}{\hbar} &=& \sqrt{\frac{E_J}{2E_C}} (2(1-\eta))^{5/4} \times \bar{S}_{\rm local}, \\
\bar{S}_{\rm local} &=& \int d\tilde{\tau} \left[ \frac12 \left(\frac{d\tilde{\psi}}{d\tilde{\tau}}\right)^2
+ \frac12 \tilde{\psi}^2 - \frac16 \tilde{\psi}^3 \right].
\end{eqnarray}

The same variable change can be done for the general action (\ref{eqn:action})
by using the expansion around the metastable state. The result is given by
\begin{eqnarray}
\bar{S}_{\rm nonlocal} &=& \int d\tilde{\tau} \frac12 \left(\frac{d\psi}{d\tilde{\tau}}\right)^2 \nonumber \\
& & \hspace{-5mm} + \int d\tilde{\tau} \int d\tilde{\tau'} \tilde{\beta}(\tilde{\tau}-\tilde{\tau}')
\left[ - \frac16 \bar{\psi}^3 + \frac12 \bar{\psi}^2 \right],
\label{eq:act3}
\end{eqnarray}
where $\tilde{\beta}(\tilde{\tau}) = \beta(\tilde{\tau})/E_J$ and
 $\bar{\psi} = (\tilde{\psi}(\tau)+\tilde{\psi}(\tau'))/2$.
The normalized action $\bar{S}$ only depends on the properties of the $\beta$-term, and
is independent of other parameters such as the current, Josephson energy and charging energy.
The value $\bar{S}$ for the bounce solution determines the prefactor in the exponent of 
the tunneling rate, and is given by $4.8$ in the local approximation.

In order to deal with the nonlocal $\beta$-term, we need the short-time cutoff $\tau_0$.
This cutoff is of order of $\hbar/\Delta_0$ in the $d$-wave junction considered in this paper,
and is much shorter than the plasma frequency $\omega_0$ in actual junctions. 
In the following discussion, we focus on the limit $\omega_0 \tau_0 \rightarrow 0$.
For the purpose of this appendix, it is sufficient to consider the following simplified kernel ($s>0$) (see also Appendix B)
\begin{eqnarray}
\tilde{\beta}(\tilde{\tau}) 
&=& \frac{A_s (\omega_0 \tau_0)^{s}}{(\omega_0 \tau_0)^{s+1} + |\tilde{\tau}|^{s+1}}, \\
A_s &=& \frac{s+1}{2\pi} \sin \left( \frac{\pi}{s+1} \right)
\end{eqnarray}
which decays in the power-law form $1/|\tau|^{s+1}$ in the long-time limit ($\tau \gg \tau_0$), by noting the sum rule $\int d\tilde{\tau} \tilde{\beta}(\tilde{\tau}) = 1$. 

First, we calculate the bounce solution of the action including non-local $\beta$ term
for the ohmic case ($s=1$) numerically.  We discretize the integral in the action (\ref{eq:act3})  as
\begin{eqnarray}
\bar{S} &\sim& \sum_i \frac12 \left( \frac{\psi_{i+1}-\psi_i}{\Delta \tau}\right)^2 
\Delta \tau \nonumber \\
& & \hspace{-12mm} + \sum_{i,j} \beta_{i,j} \! \left[ 
\frac12 \left(\frac{\psi_i + \psi_j}{2} \right)^2 
\! - \frac16 \left(\frac{\psi_i + \psi_j}{2}\right)^3 
\right] \! \Delta \tau,
\end{eqnarray}
where $\Delta \tau$ is a time slice and $\beta_{i,j} = \beta(\tau_i - \tau_j)\Delta \tau$.
The stationary condition $\delta S = 0$ is expressed by $\partial S/\partial \psi_i = 0$,
and the bounce solution is obtained by the equation
\begin{eqnarray}
0 &=& -\frac{\psi_{i+1} + \psi_{i-1} - 2\psi_i}{(\Delta \tau)^2}  \nonumber \\
&+& \sum_j \beta_{i,j} \left[ 
\left(\frac{\psi_i + \psi_j}{2}\right) - \frac12 \left(\frac{\psi_i+\psi_j}{2}\right)^2
\right].
\label{eq:stcond}
\end{eqnarray}
Here, we apply the Newton method to solve these nonlinear equations. 
For the actual calculation, we introduce a long-time cutoff $\tau_{\rm max}$, and
discretize the range $[ - \tau_{\rm max}, \tau_{\rm max} ]$ with $2N$ time slices. 
We consider the open boundary condition ($\psi_{-N} = \psi_N = 0$), and 
choose $N=400$ and $\tau_{\rm max} = 10$ ($\Delta \tau = 0.025$).

\begin{figure}[tbp]
\includegraphics[width = 7cm]{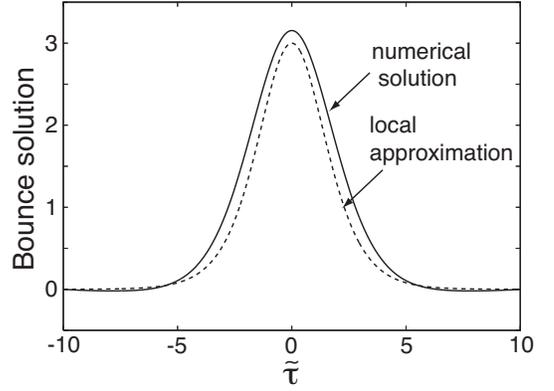}
\caption{Numerically obtained bounce solution for the action with the
ohmic non-local $\beta$ term for $\omega_0 \tau_0 = 1$ is shown by the solid line.
The usual bounce solution for the action with local approximation,
which is given by $\psi_0(\tilde{\tau}) = 3/\cosh^2(\tilde{\tau}/2)$, is also shown by the broken line.}
\label{fig:app1}
\end{figure}

We show the bounce solution for $\omega_0 \tau_0 = 1$ by the solid line
in Fig.~\ref{fig:app1}. The calculated bounce solution does not deviate so much from the usual 
bounce solution ($\psi_0(\tilde{\tau}) = 3/\cosh^2(\tilde{\tau}/2)$) for the local approximation,
which is drawn by the broken line in this figure. This result indicates that the nonlocality
of the $\beta$ term is not crucial for the shape of the bounce solution.

\begin{figure}[tbp]
\includegraphics[width=7cm]{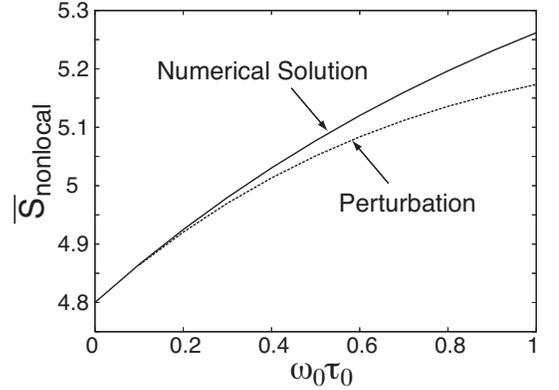}
\caption{The normalized bounce action $\bar{S}$ as a function of 
the cutoff parameter $\tau_0$ obtained by numerical calculation (solid line)
and perturbative method (dotted line). In the limit $\omega_0 \tau_0 \rightarrow 0$,
the bounce action becomes $4.8$, which corresponds to the local approximation.}
\label{fig:app2}
\end{figure}

We show the action of the bounce solution, $\bar{S}_{\mathrm{nonlocal}}$
as a function of the short-time cutoff $\tau_0$
by the solid line in Fig.~\ref{fig:app2}.
We observe that the value of the action
does not change so much from $4.8$, which corresponds to the local approximation. 
Moreover, in the limit $\omega_0 \tau_0 \rightarrow 0$, 
the calculated action approaches $4.8$. 
Similar results are obtained also for the general exponent $s$.

We can also discuss analytically the irrelevance of the nonlocality of the $\beta$ term by the perturbative
method. We introduce a parameter $\lambda$, which expresses the nonlocality of the $\beta$
term, and write the action as
\begin{eqnarray}
\bar{S}(\lambda) &=& \bar{S}_{\rm local} + \lambda \Delta \bar{S} \\
\Delta \bar{S} &=& 
\int d\tilde{\tau} d\tilde{\tau}' 
\tilde{\beta}(\tilde{\tau}-\tilde{\tau}') \left[ \frac12 \bar{\psi}^2 - \frac16 \bar{\psi}^3 \right] \nonumber \\
&-&  \int d\tilde{\tau} \left[ \frac12 \tilde{\psi}^2 - \frac16 \tilde{\psi}^3 \right].
\end{eqnarray}
The local approximation corresponds to $\lambda = 0$, while the action including the full non-local $\beta$ term is obtained for $\lambda = 1$. Here, we develop the perturbation
theory with respect to $\lambda$. Up to the first order of $\lambda$, the action is
evaluated as
\begin{equation}
\bar{S}  \simeq \bar{S}_{\rm local}[\psi_0(\tau)] + \lambda \Delta \bar{S} [\psi_0(\tau)],
\label{app:eq_act}
\end{equation}
where $\psi_0(\tilde{\tau}) = 3/\cosh^2(\tilde{\tau}/2)$ is the non-perturbed bounce solution and
$\bar{S}_{\rm local}[\psi_0(\tau)] = 4.8$.
We show this approximate action with $\lambda = 1$ for the ohmic damping by the dotted line in Fig.~\ref{fig:app2}.
As seen in this figure, the perturbation method gives a reliable result especially 
in the limit $\omega_0 \tau_0 \rightarrow 0$.

In the following, we show  results by the perturbation method, which indicates 
the irrelevance of the nonlocality of the $\beta$ term
for a general exponent $s$ by focusing on the limit $\omega_0 \tau_0 \rightarrow 0$.
We can continue calculation from (\ref{app:eq_act}) as follows,
\begin{equation}
\Delta \bar{S}[\psi_0(\tau)] = \frac{1}{16} 
\int d\tilde{\tau} \int d\tilde{\tau}' \tilde{\beta}(\tilde{\tau}-\tilde{\tau}') (\psi' - \psi)^2 (\psi' + \psi - 2),
\end{equation}
where $\psi = \psi_0(\tau)$ and $\psi' =\psi_0(\tau')$. We can calculate
the leading correction with respect to small $\omega_0 \tau_0$. For $0 < s < 2$, we  obtain
\begin{equation}
\Delta \bar{S}[\psi_0(\tau)] = \frac{A_s}{16}  (\omega_0 \tau_0)^s \gamma(s),
\end{equation}
where $\gamma(s)$ is a numerical factor, which only depends on $s$:
\begin{equation}
\gamma(s) = \int d\tilde{\tau} d\tilde{\tau}' \frac{(\psi' - \psi)^2}{(\tilde{\tau} - \tilde{\tau}')^{s+1}} (\psi' + \psi - 2).
\end{equation}
For ohmic damping ($s=1$), we obtain $\Delta S \simeq 0.084 (\omega_0 \tau_0)$.
For $s>2$, we have
\begin{equation}
\Delta \bar{S}[\psi_0(\tau)] = \frac{A_s (\omega_0 \tau_0)^2}{3} \Gamma\left(\frac{s-2}{s+1}\right)
\Gamma \left(\frac{s+4}{s+1}\right).
\end{equation}
We note that, for any $s$, the correction by the nonlocality of the $\beta$-term
disappears in the limit of $\omega_0 \tau_0 \rightarrow 0$. Therefore, 
we conclude that the nonlocal effect can be neglected at least in the calculation of the MQT rate.

Finally, we comment on nonlocal effects of the $\beta$ term in macroscopic 
quantum coherence. In this case, we have to treat the path with many instantons.
The interaction between instantons governs the coherence of the superposition
between distinct macroscopic states. In this calculation, there is a possibility that
the nonlocal effects of the $\beta$ term become important. Detailed discussion
along this line is an interesting future problem.

\section{Derivation of the asymptotic forms of $\alpha(\tau)$ and $\beta(\tau)$}\begin{widetext}
Here, we derive the asymptotic forms of $\alpha(\tau)$ and $\beta(\tau)$ for $c$-axis junctions at $\gamma=0$.\cite{rf:MQT3,Kawabata1,Joglekar,Tanaka} 
We define the order parameters as 
$\Delta_1 (\theta)= \Delta_2 (\theta)= \Delta_0 \cos 2 \theta \equiv \Delta(\theta)$. 
The dissipation kernel $\alpha(\tau)$ and the Josephson kernel $\beta(\tau)$ are defined as 
\begin{eqnarray}
\alpha (\tau )  &=&  
-\frac{2}{\hbar}
\sum_{\mbox{\boldmath $k$},\mbox{\boldmath $k$}'}
\left| t (\mbox{\boldmath $k$},\mbox{\boldmath $k$}')
  \right|^2
  {\cal G}_1 \left( \mbox{\boldmath $k$},\tau \right)
 {\cal G}_2 \left( \mbox{\boldmath $k$}',-\tau \right)
,\\
\beta (\tau ) &=& 
-\frac{2}{\hbar}
\sum_{\mbox{\boldmath $k$},\mbox{\boldmath $k$}'}
\left| t (\mbox{\boldmath $k$},\mbox{\boldmath $k$}')
\right|^2
{\cal F}_1 \left( \mbox{\boldmath $k$},\tau \right)
{\cal F}_2^\dagger \left( \mbox{\boldmath $k$}',-\tau \right).
\end{eqnarray}

Here, Matsubara Green's functions at the zero temperature are given by 
\begin{eqnarray}
{\cal G}_1 ( \mbox{\boldmath $k$},\tau)
&= &
-\frac{1}{2}
\left( 
     1
     +
      \frac{\xi_{ k} }
               {E_{ \mbox{\boldmath $k$}}}
\right)
e^{-\frac{\tau}{\hbar} E_{ \mbox{\boldmath $k$}}}
\Theta (\tau)
+
\frac{1}{2}
\left( 
     1
     -
      \frac{\xi_{ k} }
               {E_{ \mbox{\boldmath $k$}}}
\right)
e^{\frac{\tau}{\hbar} E_{ \mbox{\boldmath $k$}}}
\Theta (-\tau),
\\
{\cal G}_2 ( \mbox{\boldmath $k$}',-\tau)
&= &
\frac{1}{2}
\left( 
     1
     -
      \frac{\xi_{ k'} }
               {E_{ \mbox{\boldmath $k$}'} }
\right)
e^{-\frac{\tau}{\hbar} E_{ \mbox{\boldmath $k$}'}}
\Theta (\tau)
-
\frac{1}{2}
\left( 
     1
     +
      \frac{\xi_{ k'}}
               {E_{ \mbox{\boldmath $k$}'}}
\right)
e^{\frac{\tau}{\hbar} E_{ \mbox{\boldmath $k$}'}}
\Theta (-\tau), \\
{\cal F}_1 ( \mbox{\boldmath $k$},\tau)
&=&
\frac{\Delta_{ \mbox{\boldmath $k$}}}{2E_{ \mbox{\boldmath $k$}}}
e^{-\frac{\left| \tau \right|}{\hbar} E_{ \mbox{\boldmath $k$}}},\\
{\cal F}_2^\dagger ( \mbox{\boldmath $k$}',-\tau)
&=&
\frac{\Delta_{ \mbox{\boldmath $k$}'}}{2{E_{ \mbox{\boldmath $k$}'} }}
e^{-\frac{\left| \tau \right|}{\hbar} E_{ \mbox{\boldmath $k$}'}}
\end{eqnarray}
where $
E_{ \mbox{\boldmath $k$} }=
\sqrt{\xi_k^2 + \Delta_{ \mbox{\boldmath $k$}}^2}
$, $\Delta_{ \mbox{\boldmath $k$}}=
\Delta(\theta)$ and $\Theta$ is the step function. 

Therefore, we have 
\begin{eqnarray}
\alpha(\tau)
&=&
\frac{1}{2\hbar}
\sum_{\mbox{\boldmath $k$}}
\sum_{\mbox{\boldmath $k$}'}
\left| 
t_{ \mbox{\boldmath $k$} , \mbox{\boldmath $k$}' }
\right|^2
\left( 
     1
     +
      \frac{\xi_{ k} }
               {E_{ \mbox{\boldmath $k$}} }
\right)
\left( 
     1
     -
      \frac{\xi_{ k'} }
               {E_{ \mbox{\boldmath $k$}'} }
\right)
e^{-\frac{\tau}{\hbar}\left(  E_{ \mbox{\boldmath $k$}} + E_{ \mbox{\boldmath $k$}'} \right)}
\Theta (\tau)
\\
&+&
\frac{1}{2\hbar}
\sum_{\mbox{\boldmath $k$}}
\sum_{\mbox{\boldmath $k$}'}
\left| 
t_{ \mbox{\boldmath $k$} , \mbox{\boldmath $k$}' }
\right|^2
\left( 
     1
     -
      \frac{\xi_{ k} }
               {E_{ \mbox{\boldmath $k$}} }
\right)
\left( 
     1
     +
      \frac{\xi_{ k'} }
               {E_{ \mbox{\boldmath $k$}'} }
\right)
e^{\frac{\tau}{\hbar}\left(  E_{ \mbox{\boldmath $k$}} + E_{ \mbox{\boldmath $k$}'} \right)}
\Theta (-\tau), \\
\beta(\tau)
&=&
-\frac{1}{2\hbar}
\sum_{\mbox{\boldmath $k$}}
\sum_{\mbox{\boldmath $k$}'}
\left| 
t_{ \mbox{\boldmath $k$} , \mbox{\boldmath $k$}'}
\right|^2
\frac{\Delta_{ \mbox{\boldmath $k$}} \Delta_{ \mbox{\boldmath $k$}'
}}{ E_{ \mbox{\boldmath $k$}}  E_{ \mbox{\boldmath $k$}'}}
e^{-\frac{\left| \tau \right|}{\hbar} \left(E_{ \mbox{\boldmath $k$}}+E_{ \mbox{\boldmath $k$}'}\right)} .
\end{eqnarray}

First, we consider the coherent tunneling model where the momentum $\mbox{\boldmath $k$} $ dependence of the
tunneling matrix element is given by 
$t_{ \mbox{\boldmath $k$} , \mbox{\boldmath $k$}' }
= t
\delta_{ \mbox{\boldmath $k$}_{\Vert} , \mbox{\boldmath $k$}_{\Vert}' }. 
$ This approximation is applicable to cross-whisker junctions with a clean insulating barrier. 
 Then,  the dissipation kernel $\alpha (\tau)$ and the Josephson kernel $\beta(\tau)$ can be calculated as 
\begin{eqnarray}
\alpha(\tau)
=
\frac{N_0^2 |t|^2}{2 \pi^2 \hbar}
\int_0^{2 \pi} d \theta
\left[
\int_{0}^{\infty} d \xi 
e^{-\frac{|\tau|}{\hbar} \sqrt{\xi^2 + \Delta(\theta)^2} }
\right]^2, \\
\beta(\tau)
=
-\frac{N_0^2 |t|^2}{2 \pi^2 \hbar}
\int_0^{2 \pi} d \theta \Delta(\theta)^2 
\left[
\int_{0}^{\infty} d \xi \frac{1}{E_{ \mbox{\boldmath $k$}}}
e^{-\frac{|\tau|}{\hbar} \sqrt{\xi^2 + \Delta(\theta)^2} }
\right]^2. 
\end{eqnarray}
By  applying a formula for the modified Bessel functions $K_1$ and $K_0$,  i.e.,\begin{eqnarray}
\int_{0}^{\infty} d \xi 
e^{-\frac{|\tau|}{\hbar} \sqrt{\xi^2 + \Delta(\theta)^2} }
=|\Delta(\theta)|
K_1\left(  \frac{|\tau| |\Delta(\theta)| }{\hbar}  \right), \\
\int_{0}^{\infty} d \xi \frac{1}{E_{ \mbox{\boldmath $k$}}}
e^{-\frac{|\tau|}{\hbar} \sqrt{\xi^2 + \Delta(\theta)^2} }
=K_0\left(  \frac{|\tau| |\Delta(\theta)| }{\hbar}  \right),
\end{eqnarray}
we get Eqs. (\ref{superohmic}) and (\ref{superohmic2})~\cite{Kawabata1,rf:Bruder95,rf:Barash95}: 
\begin{eqnarray}
\alpha(\tau)
=
\frac{N_0^2 |t|^2\Delta_0^2}{2  \pi^2 \hbar}
\int_0^{2 \pi} d \theta
\cos^2 (2 \theta)
K_1\left(  \frac{|\tau|\Delta_0 }{\hbar}  |\cos 2\theta | \right)^2  \approx  \frac{{3\hbar ^2  }}{{16\pi \Delta _0^{} }}  \frac{R_Q}{R} \frac{1}{{\left| \tau  \right|^3 }}  \quad \mbox{for} \quad \tau\gg \frac{\hbar}{\Delta_0}, 
\end{eqnarray}
 and the following expression of $\beta(\tau)$: 
\begin{eqnarray}
\beta \left( \tau  \right) =  - \frac{{\left| t \right|^2 N_0^2 \Delta _0^2 }}{{2\pi ^2 \hbar }}\int_0^{2\pi } {d\theta \cos ^2 (2\theta) K_0 \left( {\frac{{\left| \tau  \right|}}{\hbar }\Delta _0 \left| {\cos 2\theta } \right|} \right)^2 }  \approx   - \frac{{\hbar ^2 R_Q }}{{16\pi \Delta _0^{} R_{} }}\frac{1}{{\left| \tau  \right|^3 }} \quad \mbox{for} \quad \tau\gg \frac{\hbar}{\Delta_0}. 
\end{eqnarray}

Next, we consider the incoherent tunneling model with a constant tunneling matrix: 
$t_{ \mbox{\boldmath $k$} , \mbox{\boldmath $k$}' } = t$, which is applicable to cross-whisker junctions with an imperfect dirty insulating barrier. 
After some calculations similar to the coherent tunneling model, we obtain \cite{Joglekar}
\begin{eqnarray}
\alpha \left( \tau  \right) = \frac{{\left| t \right|^2 N_0^2 \Delta _0^2 }}{{2\pi ^2 \hbar }}\left[ {\int_0^{2\pi } {d\theta \left| {\cos 2\theta } \right|K_1 \left( {\frac{{\left| \tau  \right|}}{\hbar }\Delta _0 \left| {\cos 2\theta } \right|} \right)} } \right]^2  
  \approx  \frac{{\hbar ^3  }}{{\pi ^2 \Delta _0^2}} \frac{R_Q}{R_1} \frac{1}{\tau ^4 } \quad \mbox{for} \quad \tau\gg \frac{\hbar}{\Delta_0}
\end{eqnarray}
where normal resistance $R_1$ is defined by 
\begin{eqnarray}
\frac{1}{{R_1 }} = \frac{{4\pi e^2 }}{\hbar }\left| t \right|^2 N_0^2. 
\end{eqnarray}
Note that the Josephson kernel $\beta \left( \tau  \right)$ disappears due to the angular averaging in the incoherent tunneling model. 

For intrinsic Josephson junctions ($c$-axis junction with $\gamma=0$), 
tunneling matrix element can be modeled as 
$
t_{ \mbox{\boldmath $k$} , \mbox{\boldmath $k$}' }
= 
t
\cos^2 \left( 2 \theta\right)
\delta_{ \mbox{\boldmath $k$}_{\Vert} , \mbox{\boldmath $k$}_{\Vert}' }
$ 
from the first-principle band-calculations,\cite{Anderson}  where the node-to-node quasiparticle tunneling is inhibited. 
Within this model, we obtain~\cite{Tanaka}
\begin{eqnarray}
\alpha(\tau)
=
\frac{N_0^2 |t|^2\Delta_0^2}{2  \pi^2 \hbar}
\int_0^{2 \pi} d \theta
\cos^4 (2 \theta)
K_1\left(  \frac{|\tau|\Delta_0 }{\hbar}  |\cos 2\theta | \right)^2 
 \approx \frac{{45\hbar ^4  }}{{128\pi  \Delta _0^3 }}  \frac{R_Q}{R_2}  \frac{1}{{\left| \tau  \right|^5 }} \quad \mbox{for} \quad \tau\gg \frac{\hbar}{\Delta_0}
\end{eqnarray}
and
\begin{eqnarray}
\beta(\tau)
=
-\frac{N_0^2 |t|^2\Delta_0^2}{2  \pi^2 \hbar}
\int_0^{2 \pi} d \theta
\cos^4 (2 \theta)
K_0\left(  \frac{|\tau|\Delta_0 }{\hbar}  |\cos 2\theta | \right)^2 
\approx - \frac{{27\hbar ^4 R_Q }}{{128\pi \Delta _0^3 R_2 }}\frac{1}{{\left| \tau  \right|^5 }}  \quad \mbox{for} \quad \tau\gg \frac{\hbar}{\Delta_0}
\end{eqnarray}
with  normal resistance $R_2$ which is defined as 
\begin{eqnarray}
\frac{1}{{R_2 }} = \frac{{ e^2 }}{\hbar }\left| t \right|^2 N_0^2. 
\end{eqnarray}

Finally, let us summarize the results of the dissipation kernel $\alpha(\tau)$ in  this Appendix in Table \ref{table1}, including the results of $s$-wave\cite{Eckern}  and  in-plane $d$-wave junctions.~\cite{Kawabata2,Fominov,Amin04} 
A remarkable feature of in-plane junctions is the emergence of ZES, which stem from the sign change of the order parameter.~\cite{Hu,TK95,KashiwayaTanaka,Lofwander} 
In this table, we also show the spectral density $J(\omega)$ which is defined by the Fourier transformation of $\alpha(\tau)$~\cite{Eckern,Simanjuntak} 
\begin{eqnarray}
\alpha(\tau)
\propto
\int_0^\infty \frac{d \omega}{2 \pi} e^{-\omega |\tau|} J(\omega)
.
\end{eqnarray}
 This table indicates that a wide variety of quantum dissipations can be realized in $d$-wave junction only by changing the junction configuration and the barrier property. 
\begin{table}[h]
\caption
{Asymptotic forms of the dissipation kernel $\alpha (\tau)$ and the spectral density $J(\omega)$ for several types of the Josephson junction (JJ).  Here, ``Exp." denotes a function of the form $\exp(-a |\tau|)$ with a constant $a$. ``Gap type" is defined by its Fourier transformation.}
\begin{center}
\begin{tabular}{cc|cc|c}
 \hline
 Symmetry & \hspace{0.5cm}   Model of JJ & \hspace{0.5cm}   Kernel $\alpha(\tau)$ & \hspace{0.5cm} Spectral density $J(\omega)$  & \hspace{0.5cm}  Refs. \\ 
 \hline
  \hline
  $s$-wave & \hspace{0.5cm} JJ without shunt resistance & \hspace{0.5cm} Exp.  & \hspace{0.5cm} Gap type & \hspace{0.5cm} \onlinecite{Eckern} \\
 $s$-wave &  \hspace{0.5cm} JJ with shunt resistance & \hspace{0.5cm} $1/\tau^2$  & \hspace{0.5cm} $\omega^{}$ & \hspace{0.5cm} \onlinecite{Eckern} \\
 \hline
 $d$-wave &  \hspace{0.5cm} $c$-axis JJ ($\gamma \ne 0$) with coherent interlayer tunneling & \hspace{0.5cm} Exp.  & \hspace{0.5cm} Gap type  & \hspace{0.5cm} \onlinecite{Kawabata1,rf:Bruder95}\\
 $d$-wave &  \hspace{0.5cm} $c$-axis JJ ($\gamma=0$) with coherent interlayer tunneling  & \hspace{0.5cm} $1/\tau^3$  & \hspace{0.5cm} $\omega^{2}$ & \hspace{0.5cm} \onlinecite{Kawabata1,rf:Bruder95,rf:Barash95} \\
 $d$-wave & \hspace{0.5cm}  $c$-axis JJ  ($\gamma=0$)  with incoherent interlayer tunneling & \hspace{0.5cm} $1/\tau^4$  & \hspace{0.5cm}  $\omega^{3}$ & \hspace{0.5cm} \onlinecite{Joglekar}\\
  $d$-wave& \hspace{0.5cm} Intrinsic JJ   & \hspace{0.5cm} $1/\tau^5$  & \hspace{0.5cm}  $\omega^{4}$ & \hspace{0.5cm} \onlinecite{Tanaka}\\
 \hline
  $d$-wave&  \hspace{0.5cm} In-plane JJ without ZES & \hspace{0.5cm} $1/\tau^3$  & \hspace{0.5cm}  $\omega^{2}$ & \hspace{0.5cm} \onlinecite{Fominov,Kawabata2}\\
 $d$-wave & \hspace{0.5cm} In-plane JJ with ZES & \hspace{0.5cm} $1/\tau^2$  & \hspace{0.5cm}  $\omega^{}$ & \hspace{0.5cm} \onlinecite{Kawabata2,Kawabata5,Amin04}
 \\ 
 $d$-wave & \hspace{0.5cm} In-plane s-wave/d-wave JJ & \hspace{0.5cm} Exp.  & \hspace{0.5cm} Gap type & \hspace{0.5cm} \onlinecite{Kawabata6-2}
 \\ 
  \hline
\end{tabular}
\end{center}
\label{table1} 
\end{table}
\end{widetext}

\end{document}